\documentclass[twocolumn,preprintnumbers,amsmath,amssymb,superscriptaddress]{revtex4}
\usepackage{graphicx}
\usepackage{dcolumn}
\usepackage{bm}
\usepackage{soul}
\usepackage{color}
\usepackage{epstopdf}
\usepackage[version=3]{mhchem}
\usepackage{mhchem}
\usepackage{lipsum}
\usepackage[outercaption]{sidecap}
\usepackage{floatrow}
\usepackage{hyperref}
\usepackage{multirow}
\begin{document}


\title{Toward a Better Understanding of the Photothermal Heating of High-Entropy-Alloy Nanoparticles}
\author{Ngo T. Que}
\affiliation{Phenikaa Institute for Advanced Study, Phenikaa University, Hanoi 12116, Vietnam}
\author{Do T. Nga}
\email{dtnga@iop.vast.vn}
\affiliation{Institute of Physics, Vietnam Academy of Science and Technology, 10 Dao Tan, Ba Dinh, Hanoi 12116, Vietnam}
\author{Anh D. Phan}
\affiliation{Faculty of Materials Science and Engineering, Phenikaa University, Hanoi 12116, Vietnam}
\affiliation{Phenikaa Institute for Advanced Study, Phenikaa University, Hanoi 12116, Vietnam}
\author{Le M. Tu}
\affiliation{Faculty of Materials Science and Engineering, Phenikaa University, Hanoi 12116, Vietnam}
\date{\today}

\begin{abstract}
We present a theoretical approach, for the first time, to investigate optical and photothermal properties of high-entropy alloy nanoparticles with a focus on FeCoNi-based alloys. We systematically analyze the absorption spectra of spherical nanoparticles composed of pure metals and alloys in various surrounding media. Through comparison with experimental data, we select appropriate dielectric data for the constituent elements to accurately compute absorption spectra for FeCoNi-based high-entropy-alloy nanoparticles. Then, we predict the temperature rise over time within a substrate comprised of Fe nanoparticles exposed to solar irradiation and find quantitative agreement with experimental data for FeCoNi nanoparticles reported in previous studies. The striking similarity between the optical and photothermal behaviors of FeCoNi nanoparticles and their pure iron counterparts suggests that iron nanoparticles can effectively serve as a model for understanding the optical and thermal response of FeCoNi-based alloy nanoparticles. These findings offer a simplified approach for theoretical modeling of complex high-entropy alloys and provide valuable insights into their nanoscale optical behavior.
\end{abstract}
\maketitle
\section{Introduction}
Alloys, materials synthesized through the combination of two or more metallic elements, are fundamental to contemporary engineering and technology, spanning infrastructure, transportation, and consumer goods \cite{41,42,33}. Alloy design has long faced a persistent trade-off where increased hardness often leads to brittleness and enhancing ductility tends to decrease strength. The recent emergence of medium- and high-entropy alloys (MEAs and HEAs), with their significant compositional disorder, offers the potential to circumvent this limitation \cite{43,44,45,46}. These novel materials exhibit a remarkable combination of hardness and ductility, surpassing the limitations of conventional alloys. 

Among many types of HEAs, FeCoNi-based high-entropy alloys have garnered considerable attention due to their exceptional combination of mechanical, physical, and chemical properties \cite{33,1,46,47}.  With iron (Fe), cobalt (Co), and nickel (Ni) in near-equimolar proportions, these alloys possess remarkable strength, ductility, hardness, and corrosion resistance. The properties are ideal for demanding environments encountered in aerospace, automotive, and marine applications \cite{33,1,46,47}. Notably, the use of abundant and cheap elements such as Fe and Al in HEAs not only reduces the dependence on Ni and Co but also maintains comparable properties, addressing concerns about resource depletion. Additionally, their adaptable magnetic properties hold promise for advancing technologies such as magnetic recording, spintronics, and magnetic refrigeration \cite{47,48,49}.  Research into FeCoNi-based HEAs deepens fundamental understandings of structure-property relationships in complex alloys and aids in the design of advanced materials for diverse industries.

Recently, it has been found that investigating optical properties and photothermal effects of high-entropy alloys is crucial due to their potential in various technological applications, and the inherent scientific challenges \cite{1,33,50,51,52,53}. Applications in optical coatings, sensors, and photonic devices necessitate a deep understanding of the materials' optical properties.  Manipulating photothermal heating could revolutionize energy conversion, photothermal therapy, and thermal management \cite{1,33,50,51,52,53}. However, the inherent structural complexity of HEAs poses challenges for theoretical predictions and precise control of their optical and photothermal properties.

Despite significant advancements in computational methods, conventional approaches for simulating photothermal effects of high-entropy alloys remain limited. Density functional theory (DFT) and molecular dynamics (MD) simulations encounter challenges in accurately describing the complex interplay among diverse elements within HEAs \cite{54,55,33}. The difficulty of accurately determining interatomic force fields in multi-component alloys historically limited the precision of modeling their electronic band structure and light-matter interactions \cite{54,55,33}. However, advancements in machine-learning interatomic potentials now allow for a more reliable description of these force fields in HEAs \cite{60,61,62,63}. Combining molecular dynamics sampling with DFT calculations enables more accurate and scalable predictions of optical and electronic properties. While machine-learning potentials can be highly accurate, their accuracy is fundamentally limited by the data they are trained on. These simulations become impractical for larger systems because the required computational resources increase rapidly with system size and this problem limits their use to smaller clusters or periodic structures. Other techniques such as the finite difference time domain (FDTD) \cite{56, 22, 23, 24}, coupled dipole approximation (CDA) or discrete dipole approximation (DDA) \cite{57, 58}, boundary element method, finite element method (FEM) \cite{22, 23, 24} used in software packages such as COMSOL or CST \cite{22,23}, and Green's dyadic method \cite{8, 9} based on solving Maxwell's equations have been employed to analyze optical spectra and electromagnetic field distribution of nanostructures in complex environments, their application to HEAs remains limited. These computational techniques depend on accurate dielectric functions of HEAs, which are challenging to determine due to the complex multi-element composition of these materials. In the absence of experimental data, it is possible to employ DFT simulations to determine the bulk dielectric function of metals and alloys, which can then serve as input for FDTD or GDM solvers to calculate the optical spectra of finite-size nanostructures \cite{64,65,66,67}. While DFT-based approaches account for atomic-level nonlinear interactions, previous studies \cite{65,66} reported quantitative discrepancies between theoretical predictions and experimental results with pronounced deviations observed at longer wavelengths. These substantial discrepancies at longer wavelengths cause inaccuracies when predicting solar photothermal heating in HEA nanostructures. Furthermore, simulating heat transfer processes caused by light absorption in HEAs, particularly for larger systems, can be computationally demanding. Therefore, novel theoretical approaches are needed to advance both fundamental materials science and the development of innovative applications in this field.

In this work, we develop the Green's dyadic method to compute the absorption spectra of both pure and alloy nanoparticles, and use this approach to study photothermal effects of FeCoNi-based high-entropy alloy nanoparticles. To ensure accurate predictions, we compare simulated absorption spectra of pure metals and alloys with experimental data to identify the most suitable dielectric data for each constituent element. We then employ two approaches for modeling alloy nanoparticles: random distribution of constituent atoms to mimic experimental structures and an effective medium approximation to estimate the dielectric function. A comparative analysis of these methods reveals the most effective approach for investigating the optical response in HEAs. Finally, we calculate the light-induced temperature rise of nanostructures under solar irradiation and compare the results with experimental photothermal data to provide insights into the underlying physical mechanisms.

\section{Theoretical background}
By utilizing the python Green's Dyadic Method (pyGDM) \cite{8,9}, we calculate the absorption spectrum of spherical nanoparticles comprising pure materials and HEAs in different environments. Nanostructures are described as a three-dimensional grid using conventional volume discretization techniques. The simulation then applies the coupled dipole approximation to analyze the geometry. Each mesh point acts as a small dipole that can oscillate in response to an incident electromagnetic field. Under the illumination of a normally incident plane wave, we compute the field within each mesh cell inside the nanostructure. Various parameters within both the near-field and far-field regions can be deduced from the internal fields acquired through CDA simulations.

Key inputs for the pyGDM simulation, apart from the geometry, include the refractive index of the environment ($n_m$) and the dielectric properties of material constituents within the structures. Each mesh point within the discretized grid is assigned a dielectric constant corresponding to a specific material. To choose suitable dielectric data for material constituents in HEAs, it is essential to compare the theoretical absorption of single-material nanostructures calculated using pyGDM with various dielectric datasets to their experimental counterparts. It is worth noting that pyGDM simulations provide the absorption cross-section of nanoparticles, $Q_{abs}$, while experimentalists measure nanoparticle absorbance. To ensure fair comparisons between simulation and experiment, we normalize the absorption data with the maximum absorption value at the optical peak.

When modeling alloys with random mixing components, once the dielectric data of each material in the alloy is known, there are two ways to compute the optical spectrum. First, different dielectric functions are randomly distributed among the mesh points to reflect the atomic-scale randomness HEAs. This random assignment captures the spatial variation in material properties across the structure as each constituent retains its characteristic bulk dielectric function. The underlying assumption is that the different components do not interact significantly to preserve their individual dielectric properties as though they were isolated. This treatment is similar to physical behaviors observed in stacked systems, where multilayers are stacked upon each other. In finite element simulations \cite{22,23,24}, interactions among material components are disregarded. In a recent work \cite{68}, the authors discussed the potential of pyGDM in binary alloy research and employed a similar approximation to that used in our work. Our random mixing model can investigate the complex electromagnetic behavior of HEAs where the randomness of atomic arrangement leads to varying local responses to light and electromagnetic fields. This approach allows the pyGDM simulation to incorporate the diverse and heterogeneous nature of HEAs while ensuring computational efficiency and realistic optical property predictions. Second, we employ the effective medium approximation to treat the alloy as a homogeneous medium with the dielectric function calculated using the Maxwell-Garnett model \cite{15,16,21}. This method assumes that the heterogeneous distribution of different material components can be averaged to derive a uniform macroscopic dielectric response. This approximation also simplifies the complex internal structure of the alloy while still accounting for the contributions of each constituent material. The Maxwell Garnett mixing formula for the dielectric function of isotropic and multicomponent mixtures, $\varepsilon_{MG}$, is
\begin{eqnarray}
\frac{\varepsilon_{MG}-\varepsilon_{h}}{\varepsilon_{MG}+2\varepsilon_{h}} = \sum_{n=1}^{N}f_n\frac{\varepsilon_n-\varepsilon_h}{\varepsilon_n+2\varepsilon_h},
\label{eq:1}
\end{eqnarray}
where $\varepsilon_{h}$ is the dielectric function of host material, $N$ is the number of inclusions in the mixtures, and $\varepsilon_n$ and $f_n$ are the dielectric function and volume fraction of inclusion $n^{th}$, respectively. The Maxwell-Garnett model offers a simple and analytical way to estimate the effective dielectric function of composites \cite{15,16,21}. It is efficient and easy to implement and provide reasonable approximations for optical properties. However, the accuracy of the Maxwell-Garnett model also diminishes when interactions among inclusions become significant and complex. 

Our developed Green's dyadic method offers distinct advantages for investigating optical properties of HEA nanoparticles compared to other methods. Mie theory predicts optical spectra of spherical particles and cannot handle complex structures \cite{15,16}. Although DFT and MD simulations provide atomic-level accuracy, they are computationally constrained to small systems and struggle to capture long-range electromagnetic interactions \cite{54,55,33}. Methods such as COMSOL, CST, and FDTD are capable of handling systems with larger sizes and arbitrary shapes \cite{56, 22, 23, 24}, they require the dielectric function of HEAs as an input, which can be challenging to obtain accurately due to their complex composition. Even with advancements in machine learning potentials facilitating MD-DFT calculations \cite{60,61,62,63} and providing a more accurate absorption function of bulk HEAs, these approaches remain computationally expensive and time-consuming for large and complex HEA systems. The accuracy of the machine-learning force field strongly depends on the training dataset. Choosing appropriate machine-learning models for HEAs necessitates a thorough evaluation of their ability to capture the intricate relationships between composition, processing, microstructure, and properties. Prediction accuracy decreases when extrapolating beyond the training data range. This is particularly problematic for HEAs due to their various and complex compositional space. The DDA method can handle light scattering in complex systems, but it also becomes computationally expensive for large structures and also requires accurate dielectric functions \cite{58,57}. In contrast, the GDM efficiently combines multiple scattering and near-field effects and this approach can study large-scale HEA structures with complex electromagnetic responses. By directly calculating electric and magnetic fields, the GDM provides crucial insights into the spatial distribution of electromagnetic energy and the interaction of light with HEA structures. This macroscopic perspective complements the atomic-level details offered by machine learning potentials and MD-DFT, which primarily focus on interatomic forces, electronic structures, and absorption spectra. Furthermore, the GDM offers a significant advantage by efficiently modeling complex interactions in multi-component materials without requiring prior knowledge of the dielectric function of alloys. Therefore, our developed GDM can serve as an indispensable tool for comprehensively analyzing the optical properties of HEAs and providing a broader view that complements existing computational approaches. 

\begin{figure}[htp]
\includegraphics[width=9cm]{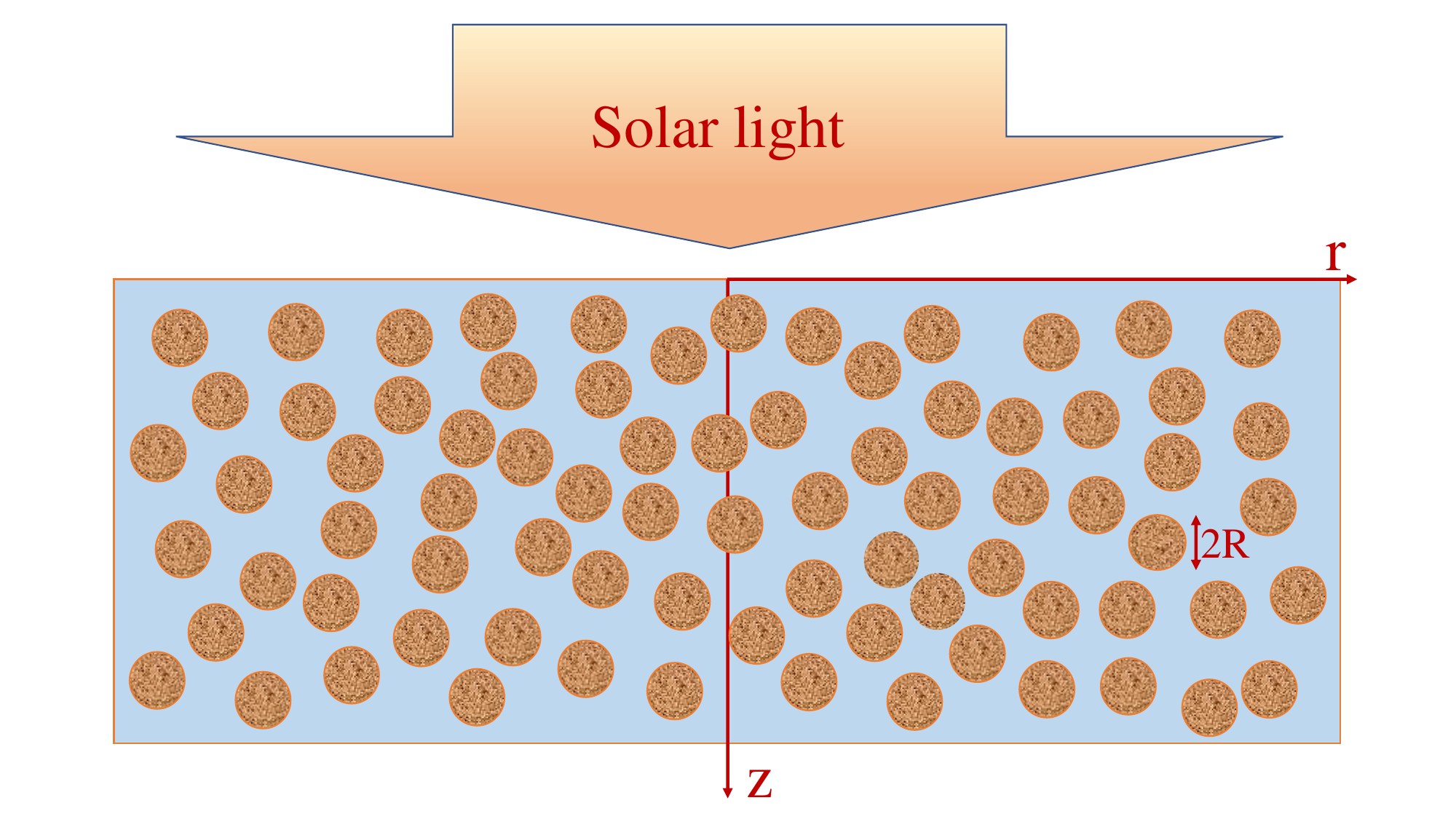}
\caption{\label{fig:0}(Color online) Illustration of a random mixture of nanoparticles with a radius of $R$ under incident solar light irradiation.}
\end{figure}

When metal nanoparticles are irradiated by the AM1.5 solar light, the nanoparticles absorb light energy, convert it into thermal energy and thereby heat the surrounding medium. Suppose that the arrangement of these nanoparticles can be modeled as a semi-infinite substrate, as illustrated in Fig. \ref{fig:0}. In prior works \cite{34,35,36,37,38}, we proposed a theoretical model based on solving the heat energy balance equation to give spatial and temporal distributions of the light-induced temperature rise, which is
\begin{widetext}
\begin{eqnarray}
\Delta T(r,z,t)=\int_{\lambda_{min}}^{\lambda_{max}}\frac{E_\lambda\alpha}{2\rho C_p}d\lambda\int_0^t \exp\left(-\frac{\beta^2r^2}{1+4\beta^2\kappa t'} \right)\frac{e^{\alpha^2\kappa t'}}{1+4\beta^2\kappa t'}\left[e^{-\alpha z}\ce{erfc}\left(\frac{2\alpha\kappa t'-z}{2\sqrt{\kappa t'}} \right)+ e^{\alpha z}\ce{erfc}\left(\frac{2\alpha\kappa t'+z}{2\sqrt{\kappa t'}} \right)\right]dt',
\label{eq:2}
\end{eqnarray}
\end{widetext}
where $E_\lambda$ is the global solar AM1.5 spectrum \cite{59}, the integral over the wavelength range in Eq. (\ref{eq:2}) ranges from $\lambda_{min}=280$ nm to $\lambda_{max}=3000$ nm, $C_p$ and $\rho$ are the specific heat capacity and mass density, respectively, $\kappa=K/\rho C_p$ is the thermal diffusivity, $K$ is the thermal conductivity, $\alpha(\lambda)=3\Phi Q_{abs}/4\pi R^3$ is the absorption coefficient at the wavelength $\lambda$ derived by assuming a random distribution of nanoparticles within the medium, and $\Phi$ is the volume fraction of nanoparticles. The parameter $\beta$ is the inverse of the light spot radius, which characterizes the spatial extent of the focused beam. Based on the specifications of the solar simulators used in Ref. \cite{1}, $1/\beta$ is set to be 30 mm. As defined in Fig. \ref{fig:0}, $z$ is the depth direction perpendicular to the substrate surface and $r$ is the radial distance parallel to the surface. 

\section{RESULTS AND DISCUSSION}
\begin{figure*}[!tp]
    \centering
\includegraphics[width=8.5cm]{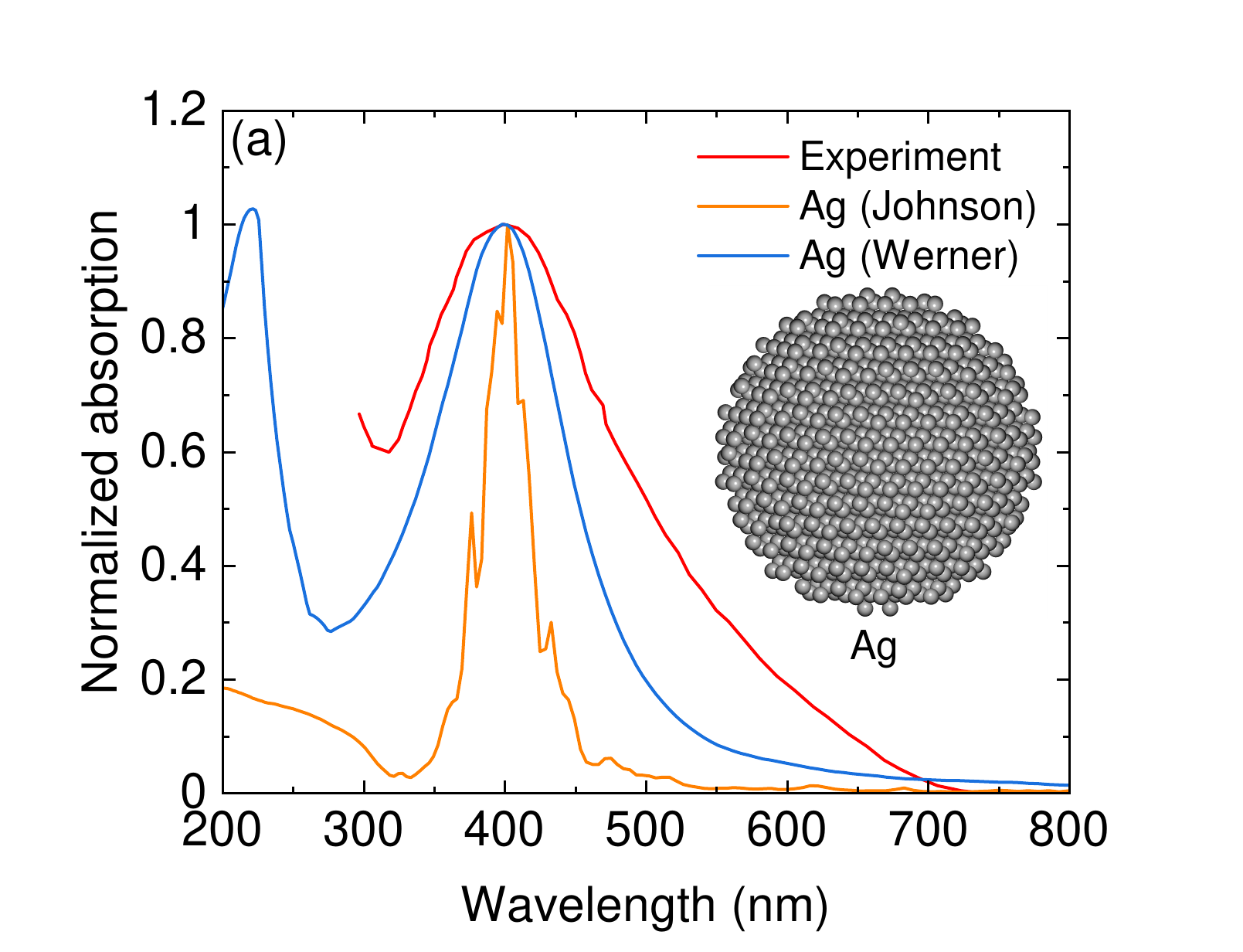}
\includegraphics[width=8.5cm]{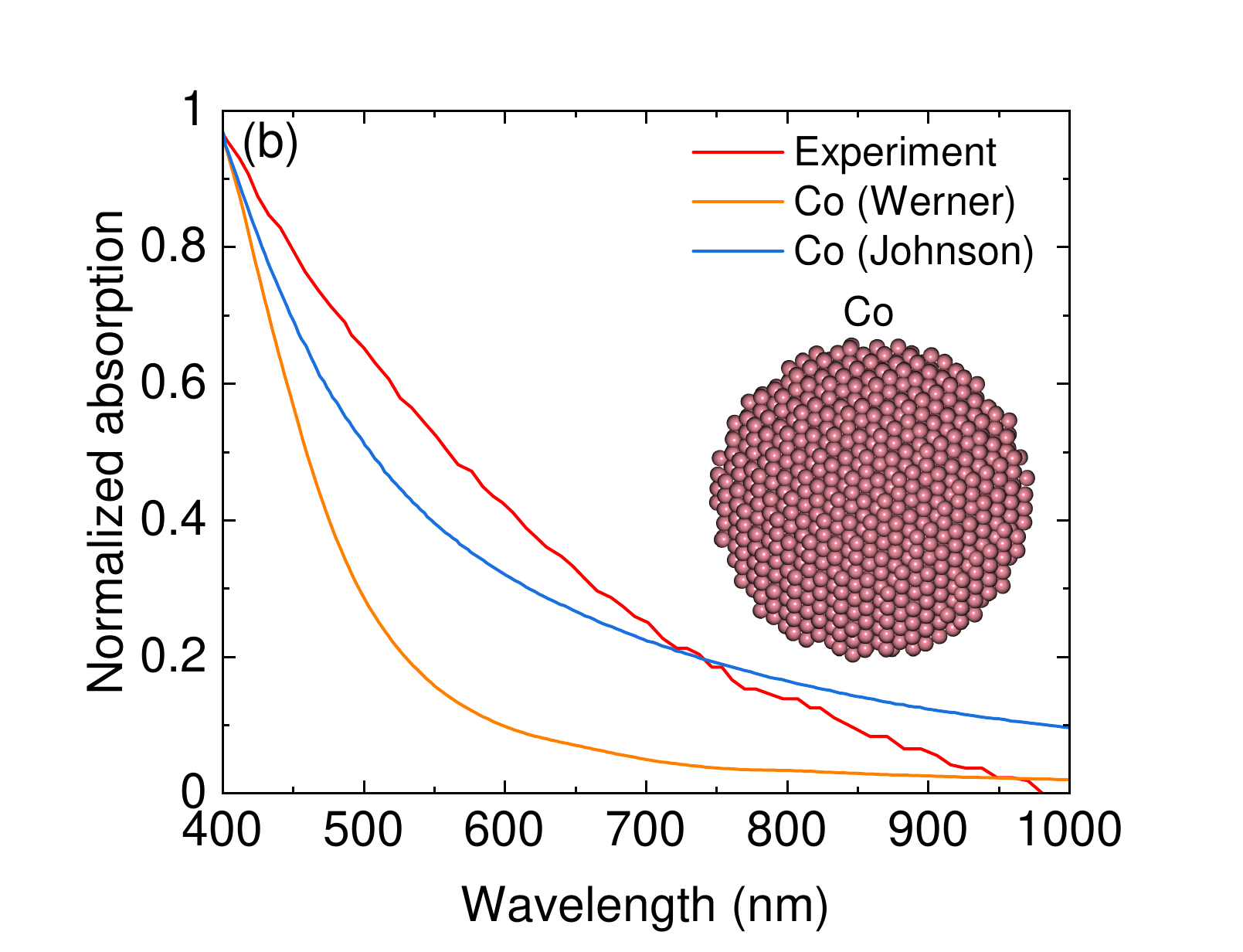}
\includegraphics[width=8.5cm]{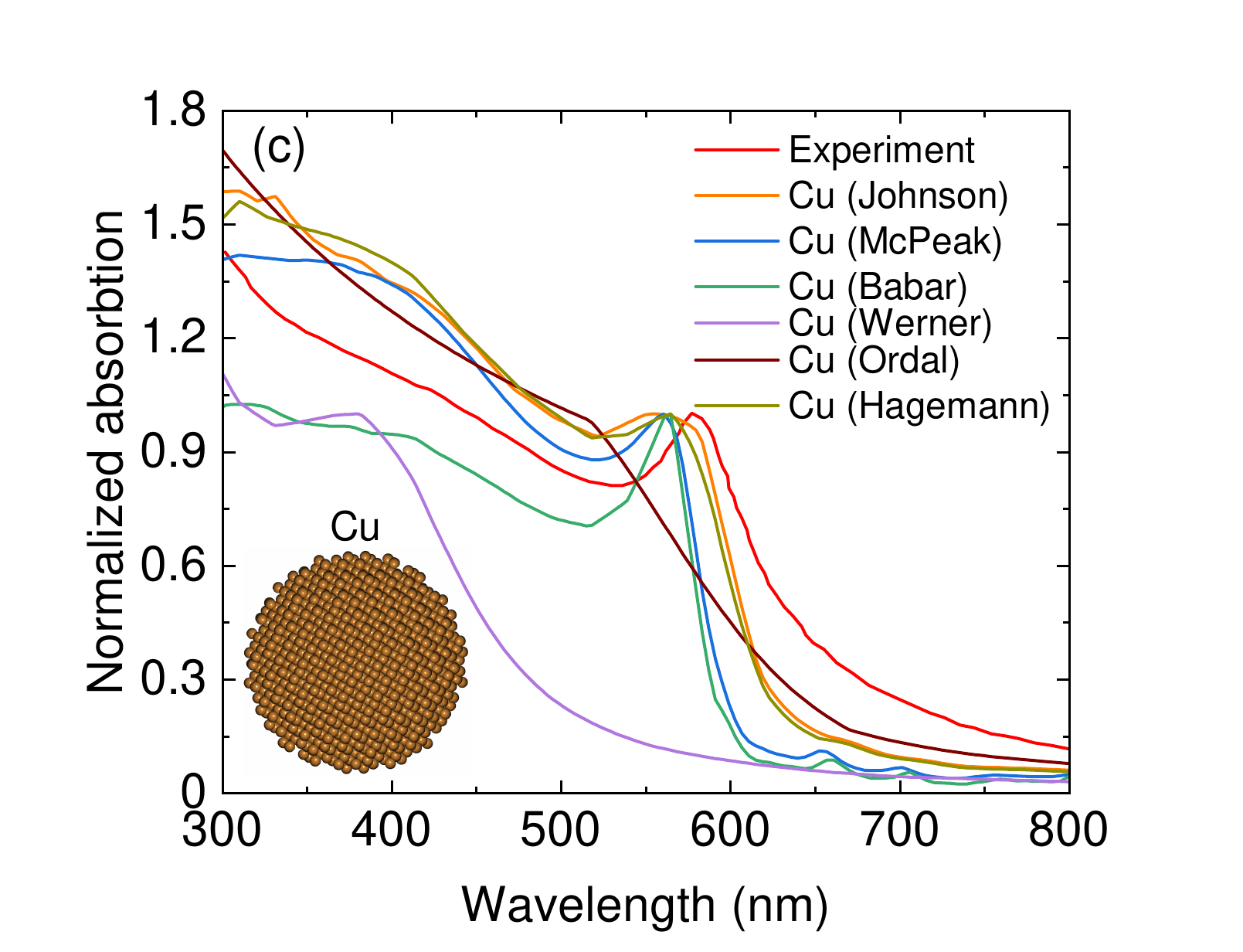}
\includegraphics[width=8.5cm]{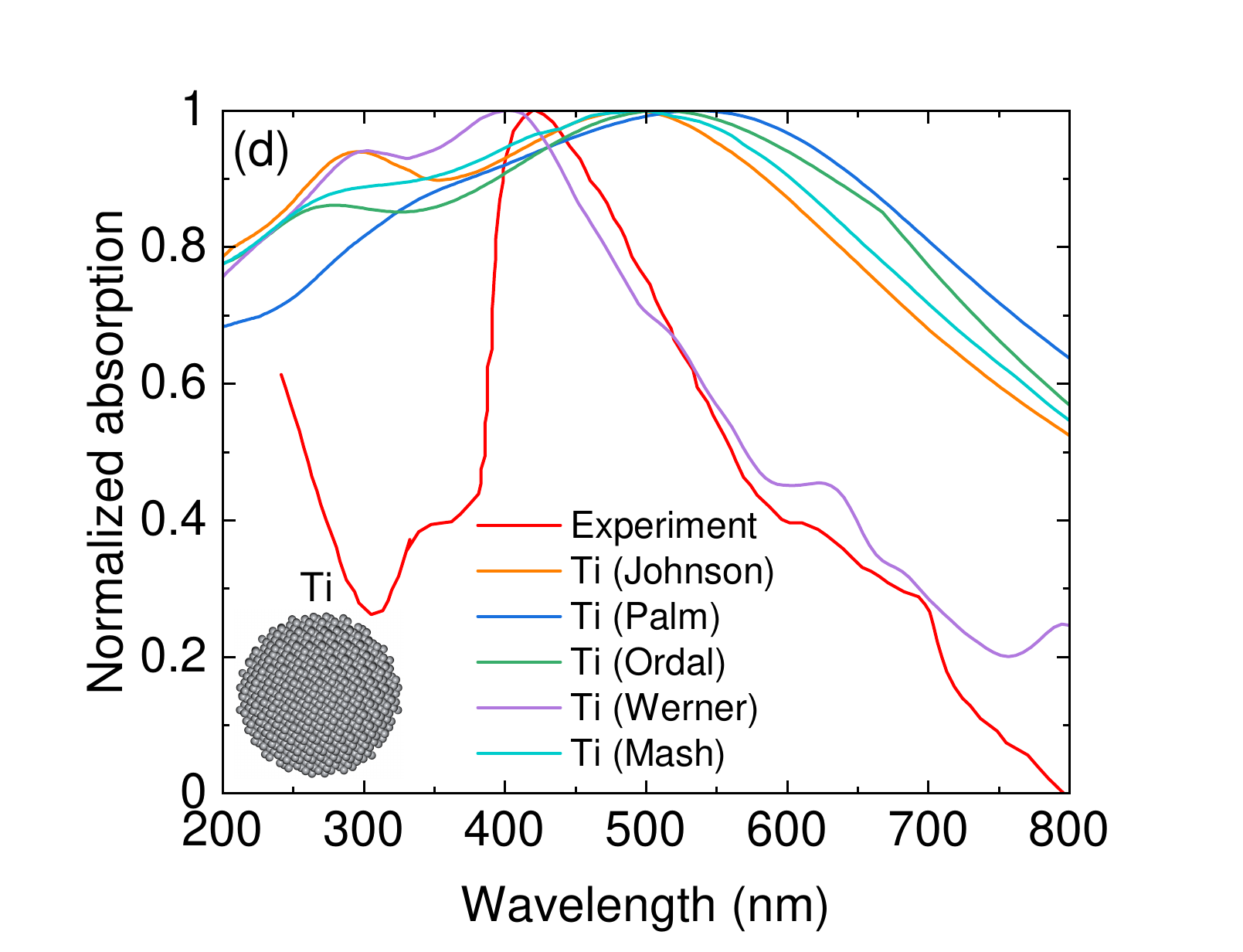}
\caption{\label{fig:1}(Color online) Normalized absorption spectra of (a) a silver nanoparticle in silica ($n_m=1.45$) with a diameter of 4 nm, (b) a cobalt nanoparticle in 1,2-dichlorobenzene ($n_m = 1.55$) with a diameter of 19 nm, (c) a copper nanoparticle in water ($n_m=1.33$) with a diameter of 2.5 nm, and (d) a titanium nanoparticle in water with a diameter of 100 nm, computed using various dielectric functions of materials. The experimental absorption data are shown in Fig. \ref{fig:1}a, \ref{fig:1}b, \ref{fig:1}c, and \ref{fig:1}d, are taken from Ref. \cite{2}, \cite{5}, \cite{6}, and \cite{7}, respectively.}
\end{figure*}

Figure \ref{fig:1}a shows experimental and theoretical normalized absorption cross section of pure silver nanoparticle embedded in silica. The theoretical absorption is numerically computed using pyGDM and two different dielectric datasets given by Johnson and Christy \cite{3} and Werner \cite{4}. The computational optical spectra are narrower than the experimental counterpart because simulations assume uniform particle size, whereas synthesized nanoparticles have a size distribution. However, it is clear to see that calculations with Werner's dielectric data provide a good quantitative agreement with experimental data \cite{2}. Therefore, in subsequent calculations, we consistently utilize the dielectric data of Werner for silver components in nanostructures.

Similarly, experimental and computational absorption spectra of other pure-material nanoparticles are compared in Fig. \ref{fig:1}b, \ref{fig:1}c, and \ref{fig:1}d to determine the most suitable dielectric data for subsequent calculations for HEAs. The pyGDM calculations for the 19-nm Co nanoparticle, using the dielectric data of Johnson and Christy \cite{3}, have a better agreement with experimental results \cite{5} compared to those employing the dielectric function of Werner \emph{et al.} \cite{4} (see Fig. \ref{fig:1}b). Furthermore, when compared to the dielectric data of McPeak \emph{et al.} \cite{25}, Babar and Weaver \cite{26}, Werner \emph{et al.} \cite{4}, and Ordal \emph{et al.} \cite{27}, Hagemann \emph{et al.} \cite{28}, integrating the dielectric dataset of Johnson and Christy \cite{3} into pyGDM simulations provide the best quantitative accordance with experimental data for the absorption spectrum of Cu nanoparticles as shown in Fig. \ref{fig:1}c. For Ti nanoparticle absorption (Fig. \ref{fig:1}d), all theoretical calculations underestimate the optical properties in the UV range. However, pyGDM calculations using the Werner dataset show the closest agreement with experimental results compared to other datasets from Johnson and Christy \cite{3}, Palm \emph{et al.} \cite{29}, Ordal \emph{et al.} \cite{30}, and Mash and Motulevich \cite{31}.

\begin{figure}[htp]
\includegraphics[width=9cm]{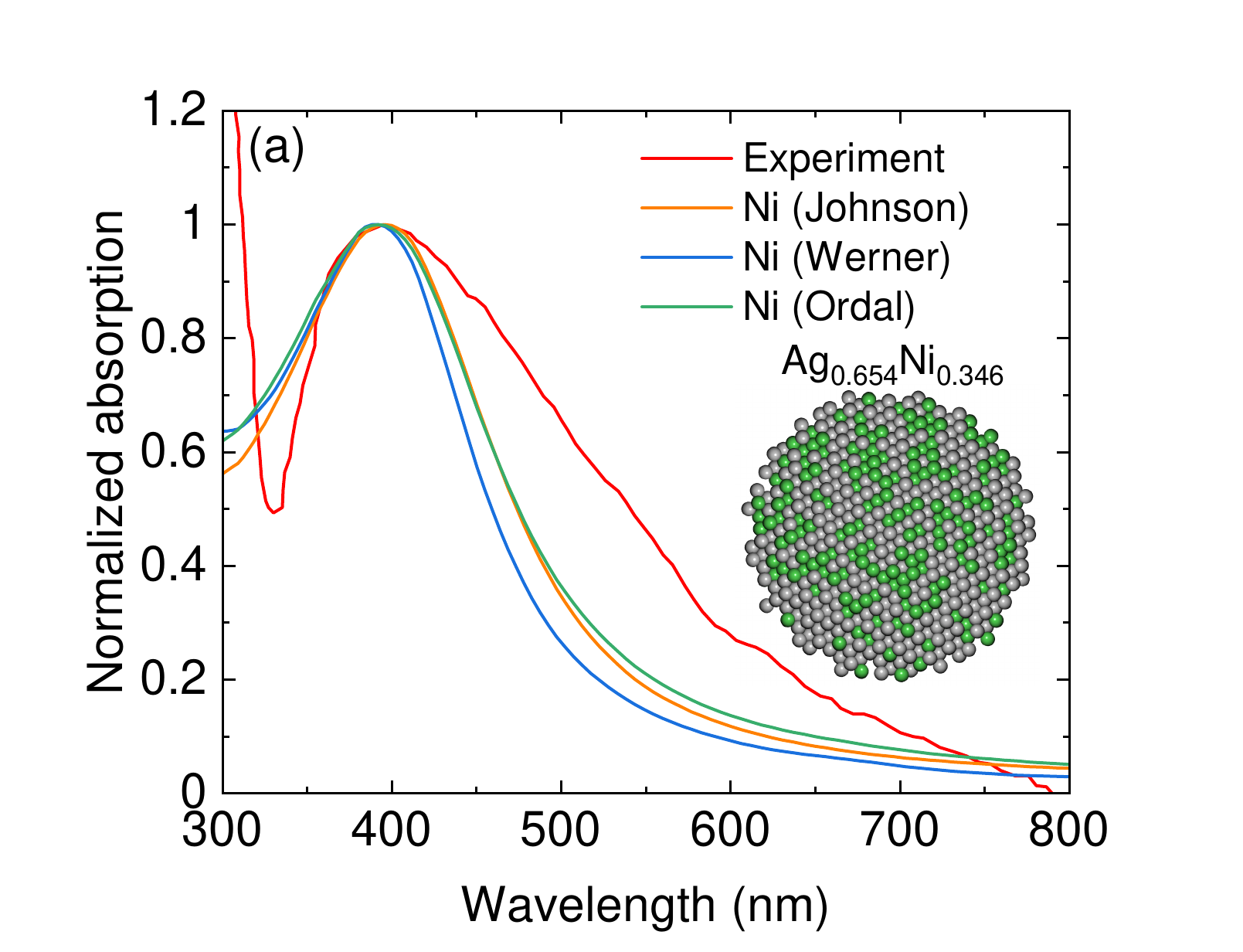}
\includegraphics[width=9cm]{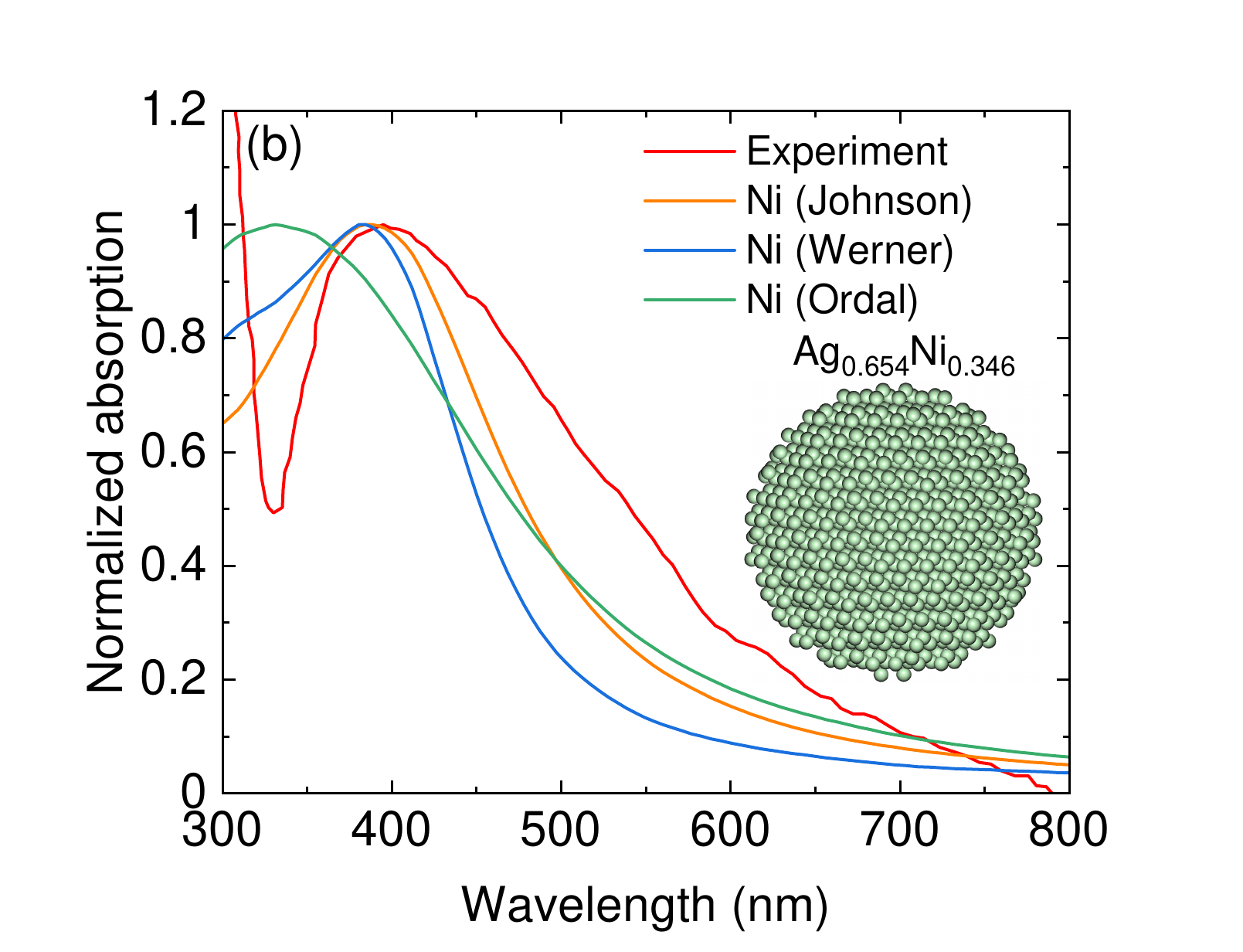}
\caption{\label{fig:2}(Color online) Normalized absorption spectrum of a \ce{Ag_{0.654}Ni_{0.346}} nanoparticle with a diameter of 38 nm in water calculated using different dielectric data of nickel reported by Johnson and Christy \cite{3}, Werner \emph{et al.}\cite{4}, and Ordal \emph{et al.} \cite{10}. The optical properties of nanoparticles are calculated using (a) the random mixing model of Ag and Ni atoms in the alloys and (b) the Maxwell-Garnett model \cite{15,16} for the dielectric function of the binary mixture. Experimental data of the absorption spectrum is taken from Ref. \cite{18}.}
\end{figure}

Given the difficulty in finding the experimental absorption spectrum of pure nickel nanoparticles, we conduct simulations to predict the absorption spectrum of \ce{Ag_{0.654}Ni_{0.346}} nanoparticles and compare computational results with experimental data \cite{18}. To achieve this, the random mixing model is employed to describe the elemental distribution within the alloy nanoparticle. Various dielectric datasets for nickel \cite{3,4,10} are then integrated into the pyGDM simulations to calculate the absorption spectrum of the alloy nanoparticle. A comparison between pyGDM simulations using various dielectric datasets and experimental results (Fig. \ref{fig:2}a) reveals that the Johnson and Christy data provides the closest match. Theoretical curves closely overlap each other and the experimental curve in the vicinity of the optical peak.

To exam the validity of the Maxwell-Garnett model in approximate optical properties of alloys, we carry out pyGDM simulations for the \ce{Ag_{0.654}Ni_{0.346}} nanoparticle with the effective dielectric function given by Eq. (\ref{eq:1}). As seen in Fig. \ref{fig:2}b, although the Maxwell-Garnett model gives a relatively good prediction for the absorption of the alloy nanoparticle, its accuracy is still less than the random mixing model. Unlike calculations in Figure 2a, theoretical curves corresponding to the use of different dielectric data for nickel exhibit significant disparities and cannot accurately predict the peak position of the experimental absorption spectrum. Clearly, this is an inherent issue of the computational model, rather than a problem with the selection of dielectric data for the material.

\begin{figure}[htp]
\includegraphics[width=9cm]{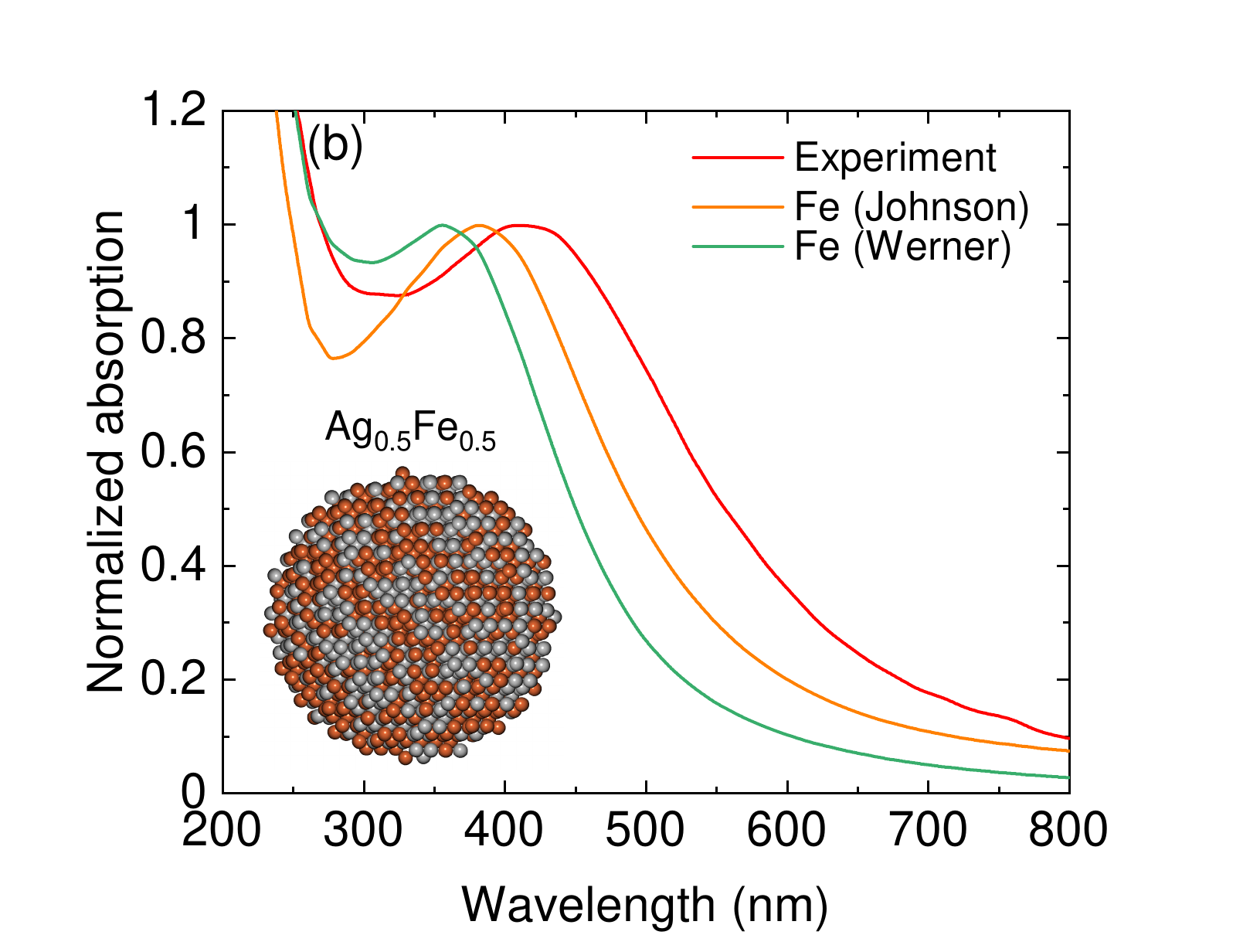}
\includegraphics[width=9cm]{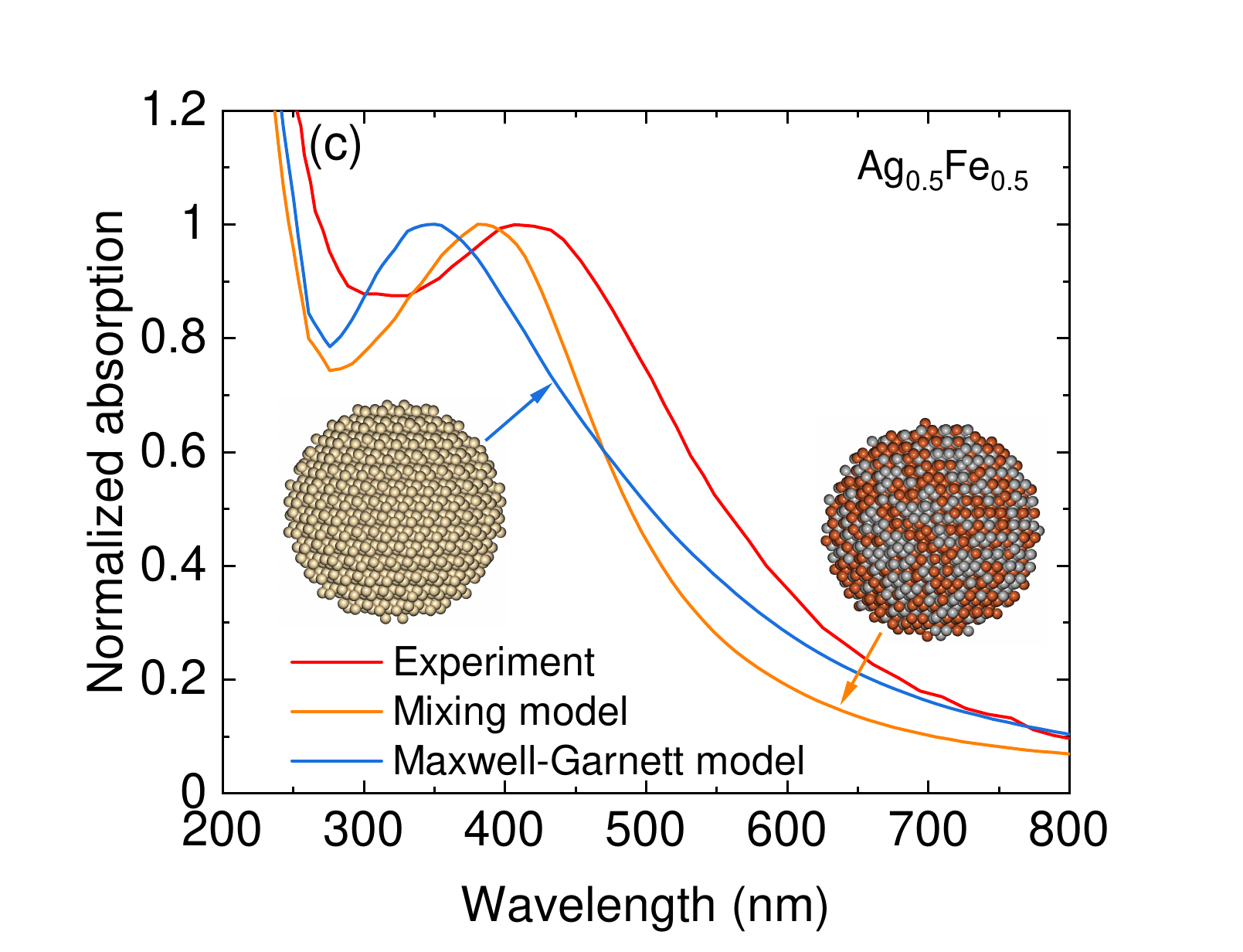}
\caption{\label{fig:3}(Color online) (a) Normalized absorption spectrum of a \ce{Ag_{0.5}Fe_{0.5}} nanoparticle with a diameter of 4.5 nm embedded in silica ($n_m=1.45$) calculated using different dielectric data of iron reported by Johnson and Christy \cite{3} and Werner \emph{et al.} \cite{4}. (b) The optical properties of nanoparticles are calculated using a random mixing model of Ag and Fe atoms in the alloy and the Maxwell-Garnett model \cite{15,16} for the dielectric function of the binary mixture. Experimental data of the absorption spectrum is taken from Ref. \cite{2}.}
\end{figure}

The same issue also arises when studying the dielectric function of iron. Due to the ease of iron oxidation during fabrication processes, it is hard to obtain the absorption spectrum of pure iron nanoparticles. Therefore, we have to investigate indirectly through the optical absorption spectrum of iron-containing alloys, in this case,  \ce{Ag_{0.5}Fe_{0.5}} nanoparticles to align with experiments in Ref. \cite{2}. Simulations with the random mixing model of Fe and Ag atoms in \ce{Ag_{0.5}Fe_{0.5}} alloy nanoparticles are carried out and contrasted with its experimental data in Fig. \ref{fig:3}a. Our theoretical calculations relatively well describe the shape and variation of the absorption spectrum as a function of the incident wavelength. Note that Ramade and his workers \cite{2} also attempted to simulate the absorption spectrum of AgFe nanoalloy systems using concentric and eccentric core-shell models. However, this model does not mimic closely experimental fabrication and leads to relatively significant discrepancies between computational and experimental results. Meanwhile, our simulations give a much closer agreement with experimental data compared to Ramade's approach. Additionally, as seen in Fig. \ref{fig:3}a, pyGDM calculations using the dielectric data of Johnson and Christy for iron \cite{3} offer the most accurate description of the optical spectrum of AgFe alloy nanoparticles. Furthermore, results in Figure 3b reaffirms the superiority of the random mixing model over the Maxwell-Garnett model when integrated into pyGDM simulations for investigating the optical properties of metallic alloys. Thus, hereafter, we only use the random mixing model for calculating the absorption of HEA nanoparticles.

\begin{figure}[htp]
\includegraphics[width=9cm]{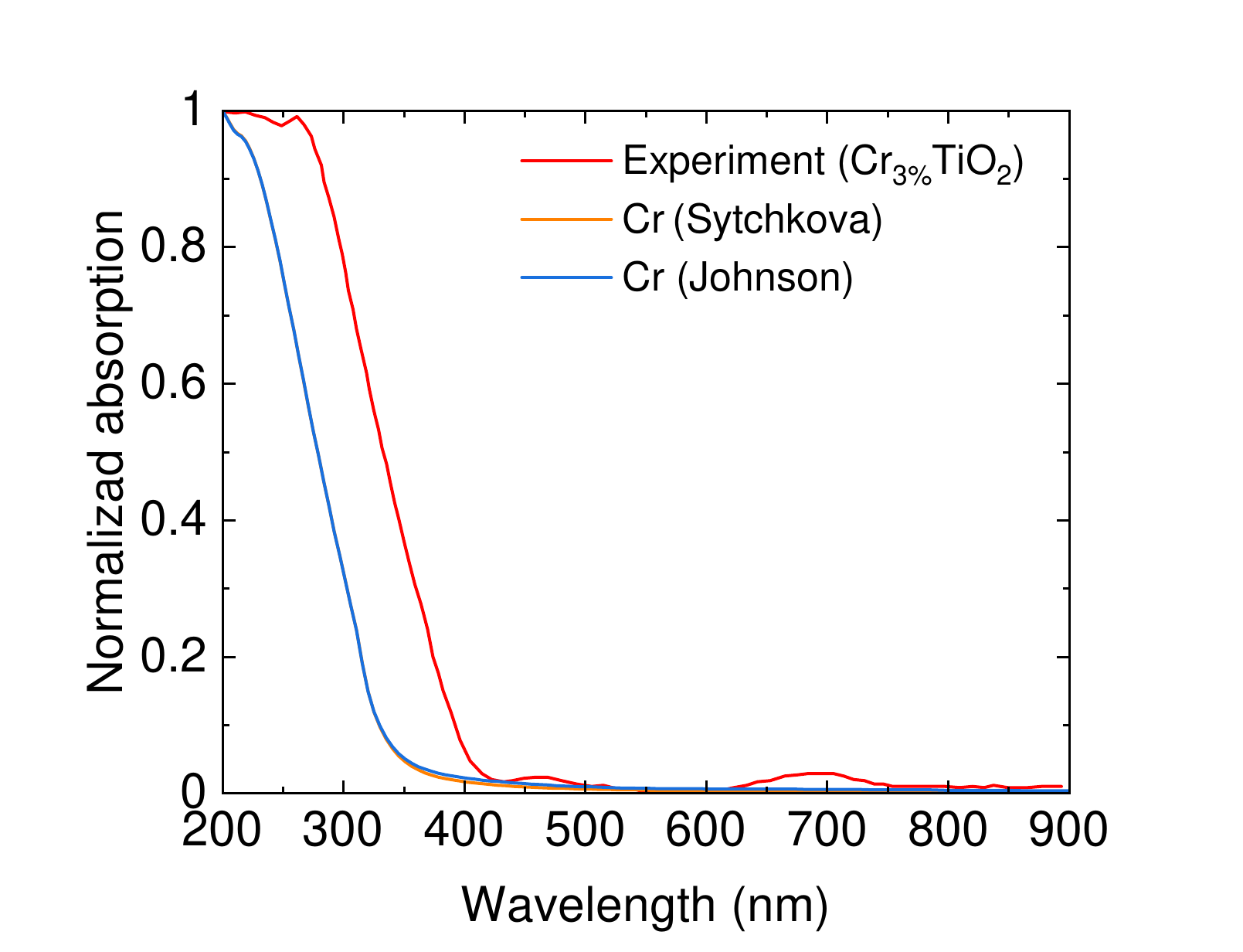}
\caption{\label{fig:4}(Color online) Normalized absorption spectrum of a 32-nm \ce{TiO_{2}} nanoparticle doped by 3$\%$ chromium in water calculated using the dielectric data of Cr from the work of Johnson and Chirsty \cite{3} and Sytchkova \emph{et al.} \cite{11}. The experimental absorption spectrum data is obtained from Ref. \cite{17}.}
\end{figure}

Figure \ref{fig:4} shows the experimental and computational absorption of 3$\%$-Cr-doped \ce{TiO_2} nanoparticle in water. Here, we use the dielectric data for \ce{TiO_2} taken from the study of Zhukovsky \emph{et al.} \cite{14}. One can also employ datasets provided by Sarkar \emph{et al.} \cite{12}, Siefke \emph{et al.} \cite{13}, or Jolivet \emph{et al.} \cite{32} for \ce{TiO_2}. Our calculations, although not shown here, indicate that the absorption spectrum has slight variation across different selections of \ce{TiO_2} dielectric data. This is similar to numerical results illustrated in Fig. \ref{fig:4}. When the optical properties of pure \ce{TiO_2} remain unchanged and only the dielectric data of Cr is varied, computational outcomes exhibit minimal alteration. The simulated data also adequately describe the shape of the experimental curve. Therefore, we employ the dielectric data provided by Johnson and Christy for Cr in subsequent calculations.

\begin{figure}[htp]
\includegraphics[width=9cm]{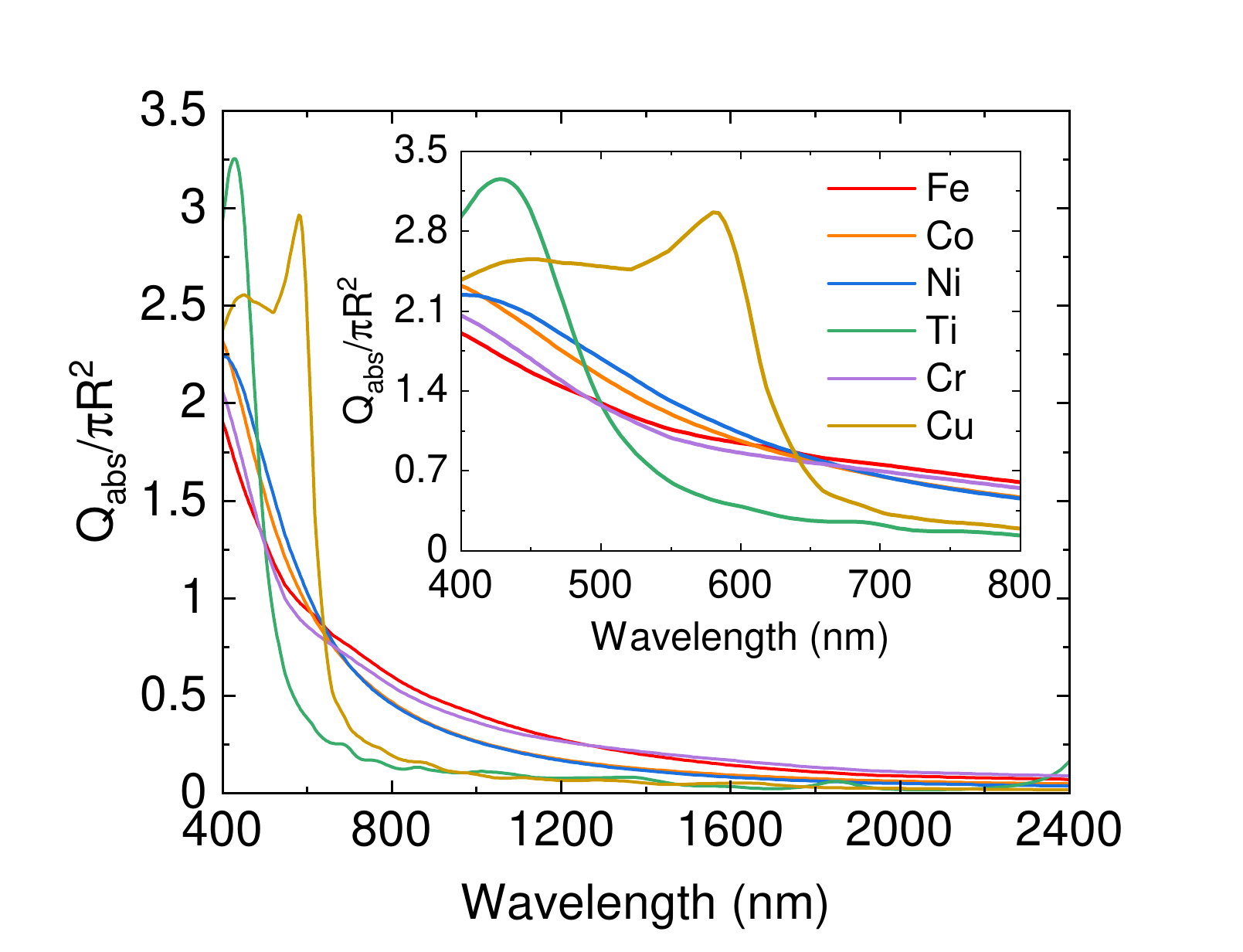}
\caption{\label{fig:5}(Color online) The theoretical absorption cross section of different pure nanoparticles with a radius $R = 30$ nm in polymer medium ($n_m = 1.4$) normalized by $\pi R^2$. The inset shows the same data as the mainframe but focuses on the wavelength range from 400 to 800 nm.}
\end{figure}

\begin{figure}[htp]
\includegraphics[width=9cm]{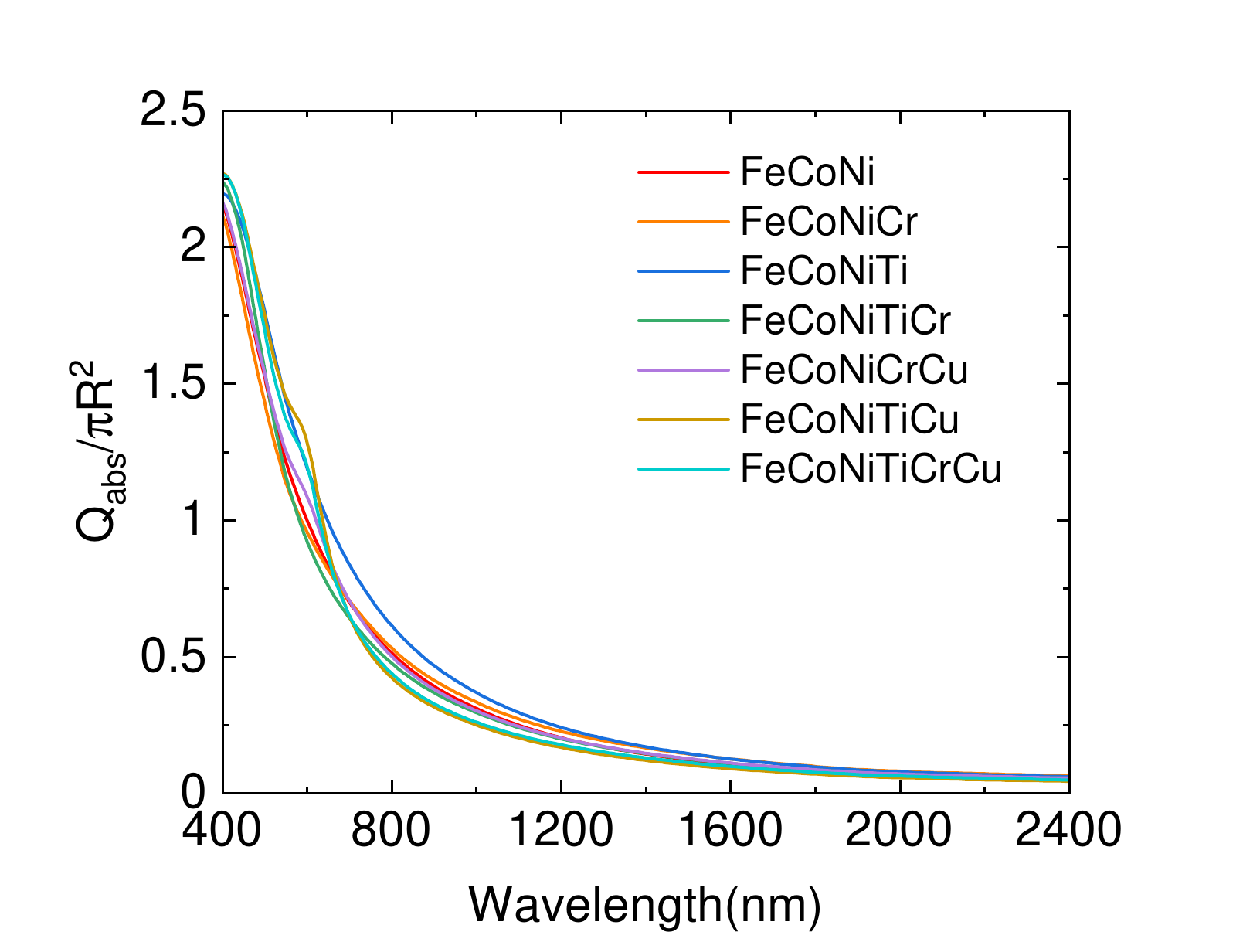}
\caption{\label{fig:6}(Color online) The theoretical absorption cross section of different FeCoNi-based alloys with a radius $R = 30$ nm normalized by $\pi R^2$.}
\end{figure}

Now, we identify the most appropriate dielectric data for Fe, Ni, Co, Cr, Ti, and Cu, which are summarized in Table \ref{table:1}, to use in hereafter pyGDM simulations. Before utilizing these data to investigate the optical properties of HEAs, we examine the absorption spectra of single-material nanoparticles with the same radius, $R = 30$ nm, embedded in polymer. Computational spectra across the solar light range are presented in Fig. \ref{fig:5}. The solar energy absorption is remarkably similar across various nanostructures except for those composed of Cu and Ti. The curves of $Q_{abs}/\pi R^2$ of Fe, Co, Ni, and Cr relatively overlap each other. This finding is completely consistent with experiments in Ref. \cite{1} and implies that absorption properties of FeCoNi-based alloys are nearly identical to those of their single-component counterparts. To validate this hypothesis, we conduct pyGDM simulations to obtain the normalized absorption spectra of FeCoNi, FeCoNiCr, FeCoNiTi, FeCoNiTiCr, FeCoNiCrCu, FeCoNiTiCu, and FeCoNiTiCrCu nanoparticles embedded in a polymer medium. The numerical results are shown in Figure \ref{fig:6}. Intriguingly, the theoretical curves exhibit substantial overlap and this result indicates a high degree of similarity in their absorption properties. Our calculations are consistent with experimental findings reported in Ref. \cite{1,33}.

\begin{table}[h]
\begin{center}
\begin{tabular}{|p{1.5cm}|p{5.5cm}| }
 \hline
 Material  & Dielectric data \\
 \hline
 Ag & Werner's data \cite{4} \\ 
 \hline
 Co  & Johnson and Christy's data \cite{3} \\
 \hline
 Cu & Johnson and Christy's data \cite{3} \\
 \hline
 Ti & Werner's data \cite{4} \\
 \hline
 Ni & Johnson and Christy's data \cite{3} \\
 \hline
 Fe & Johnson and Christy's data \cite{3} \\
 \hline
 Cr & Johnson and Christy's data \cite{3} \\
 \hline
\end{tabular}
\caption{\label{table:1} Summary of constituent elements and dielectric data used in pyGDM with the random mixing model of atoms to determine optical properties of alloys/mixtures.}
\end{center}
\end{table}

In the inset of Figure \ref{fig:5}, the surface plasmon resonance of the pure Cu nanoparticle is observed around 580 nm. However, in metallic alloys comprising various metals, the collective electron oscillations in the copper component can be significantly dampened, leading to the absence of its distinct optical peaks. Only a minor shoulder near the 600 nm wavelength is discernible in the absorption spectrum of certain alloy nanoparticles containing Cu, as presented in Figure \ref{fig:6}. This phenomenon arises primarily due to increased electron scattering caused by the alloying process. The presence of multiple elements disrupts the ordered lattice structure of the pure metal, creating a heterogeneous environment with varying electron densities and scattering centers. Consequently, the coherent oscillation of electrons is hindered and plasmon lifetime is decreased. Additionally, the alloying process can modify the electronic bandstructure of the constituent metals, leading to changes in the density of states at the Fermi level, further impacting the plasmonic response. These effects collectively contribute to the diminished intensity observed in the absorption spectra of metallic alloys compared to their pure metal counterparts. Since our calculations do not account for modifications to the electronic band structure, the latter factor is not the primary reason.

\begin{figure}[htp]
\includegraphics[width=9cm]{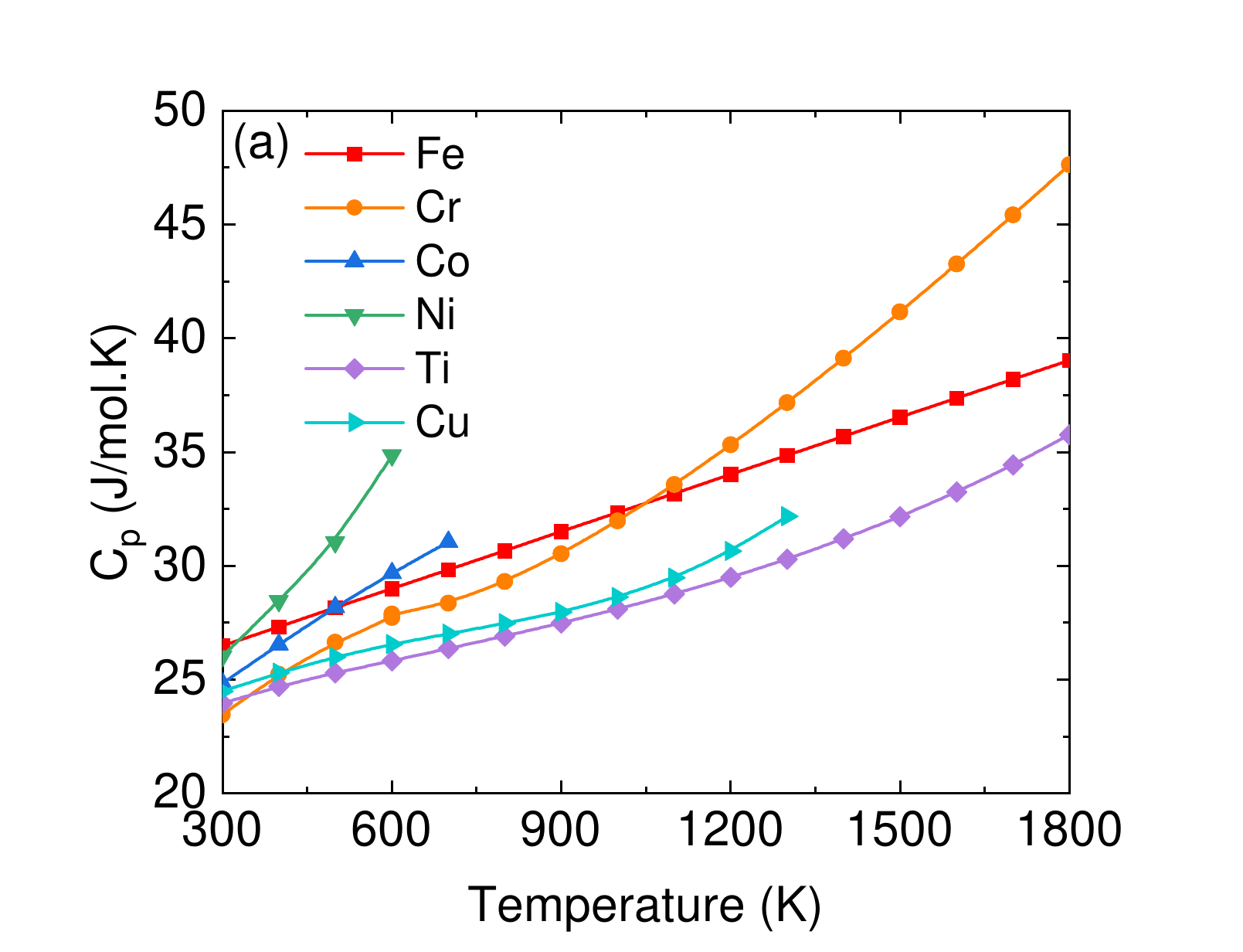}
\includegraphics[width=9cm]{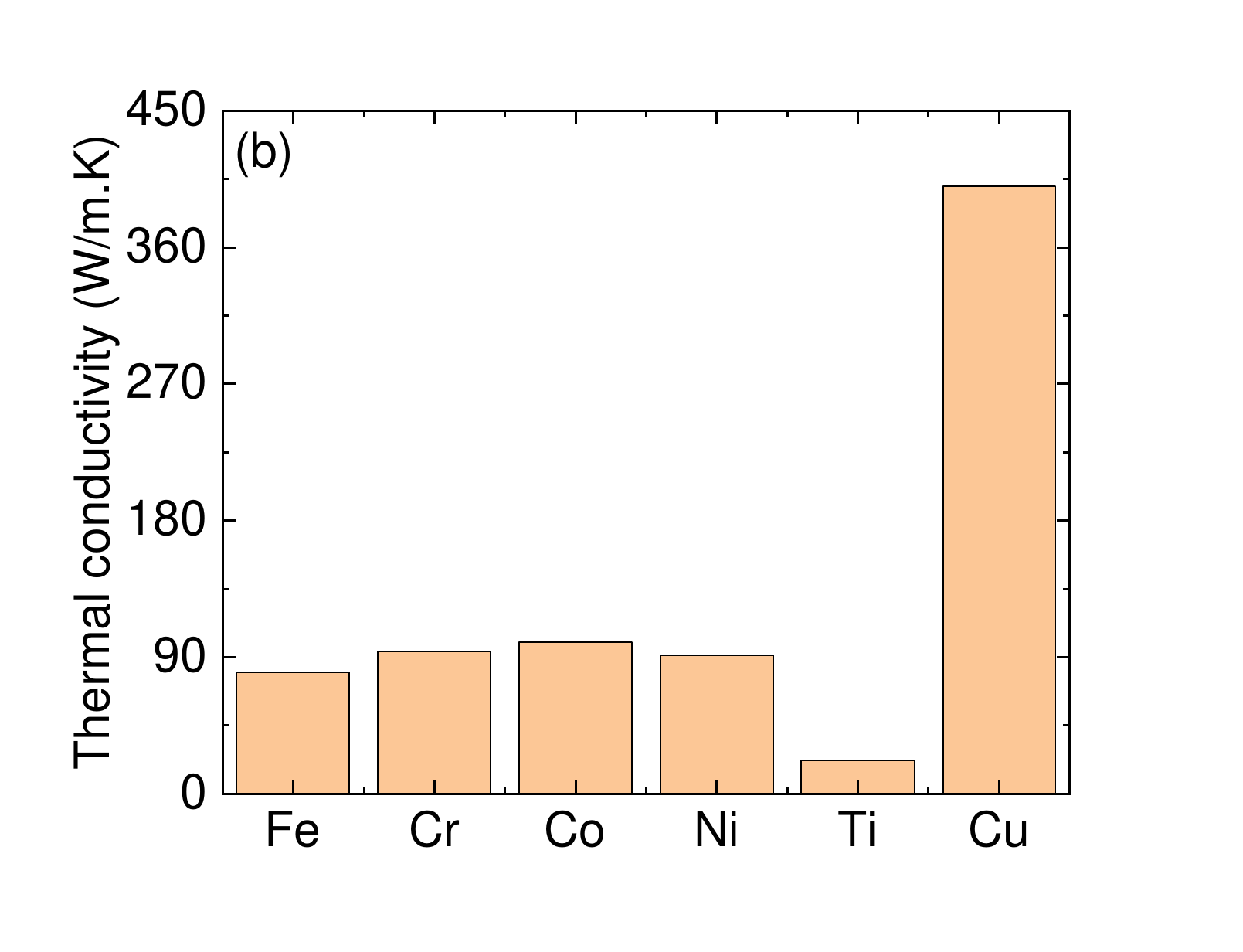}
\includegraphics[width=9cm]{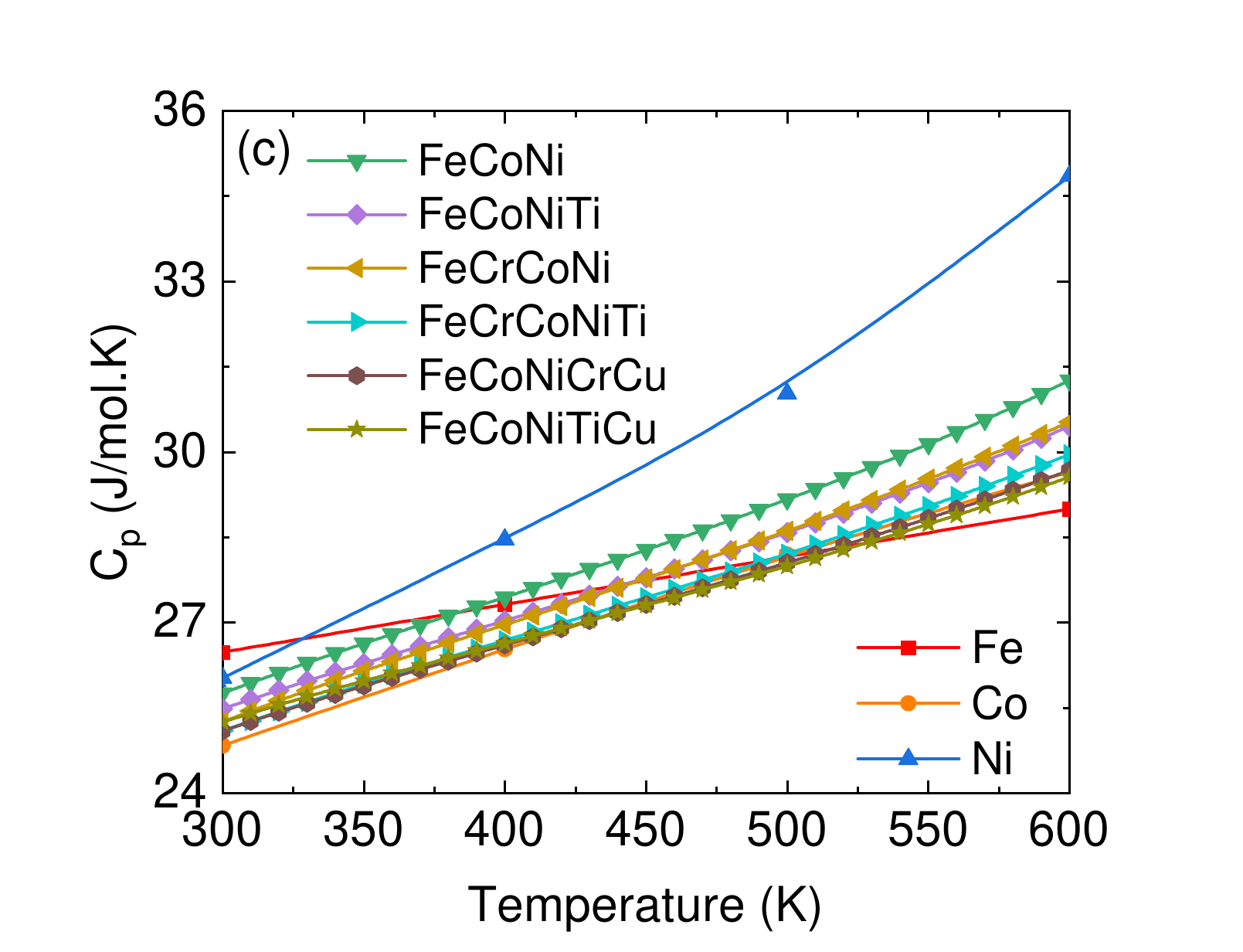}
\caption{\label{fig:7}(Color online) (a) The temperature dependence of the experimental specific heat capacity \cite{19} and (b) The thermal conductivity at room temperature \cite{20} of Fe, Cr, Co, Ni, Ti, and Cu. (c) Comparison of experimental specific heat capacity ($C_p(T)$) data for Fe, Co, and Ni \cite{19} with theoretical predictions for FeCoNi-based HEAs.}
\end{figure}

High-entropy alloy nanoparticles, when exposed to solar or laser irradiation, absorb light energy. This absorbed energy is subsequently converted into thermal energy and raises the temperature of the nanostructures through photo-to-thermal conversion \cite{1,33}. The photothermal effect of nanostructures not only depends on their absorption spectrum, but also the specific heat capacity, thermal conductivity, and mass density \cite{34,35,36,37,38}. Given the relatively similar specific heat capacity and thermal conductivity of Fe, Cr, Co, and Ni (see Figs. \ref{fig:7}a and \ref{fig:7}b), we assume these properties remain nearly constant during alloying to form metallic compounds. To validate this assumption, we calculate the effective specific heat capacity of FeCoNi-based HEAs by using the analytical expression \cite{27}
\begin{eqnarray}
C_{p,eff}(T) &=& \frac{\sum_i C_{p,i}(T)\rho_{i}\delta_i}{\sum_i\rho_{i}\delta_i},
\label{eq:5}
\end{eqnarray}
where $i$ is an element in the alloy, and $C_{p,i}$ and $\delta_i$ is the specific heat capacity and the molar fraction of element $i$, respectively. Equation (\ref{eq:5}) is derived under the following assumptions: (1) no interactions occur between components, (2) the total volume of mixture equals the sum of individual component volumes, and (3) no phase transitions occur within the studied temperature range. These assumptions appear valid for investigating the photothermal heating of HEAs.

Figure \ref{fig:7}c compares the experimental specific heat capacity $C_p(T)$ of pure Fe, Co, and Ni with the theoretical $C_p(T)$ of various FeCoNi-based HEAs calculated using Eq. (\ref{eq:5}). In high-entropy alloys, constituent elements typically exhibit near-equimolar fractions ($\delta_i$), the effective specific heat capacity $C_{p,eff}(T)$ is primarily governed by the specific heat capacities and mass densities of the individual elements. Within the temperature range of 300 $K$ to 500 $K$, the specific heat capacities of FeCoNi-based HEAs vary slightly with a difference not exceeding approximately 1 J/mol.K. The comparable thermal and optical properties of these materials suggest similar photothermal effects under identical conditions. This analysis clearly explains experimental findings observed in previous studies \cite{1,33}, wherein the temperature increases of FeCoNi-based HEAs under solar light illumination exhibit similar variations over time.

Numerical results in Fig. \ref{fig:7}c indicate that the effective specific heat capacity of FeCoNi-based HEAs is relatively close to that of pure Fe. This suggests that although HEAs possess superior mechanical properties compared to iron, their optical properties and photothermal conversion are similar. To confirm the similarity in photothermal effects, we calculate the time-dependent temperature rise at the hottest spot of the substrate containing Fe nanoparticles by using Eq. (\ref{eq:2}) and compare numerical results with photothermal experimental counterparts of FeCoNi nanoparticles in Ref. \cite{1}. 

Since the FeCoNi nanoparticles in the experiment conducted by Ref. \cite{1} were randomly dispersed in an air medium, we adopt the same description for our spherical Fe-nanoparticle systems. To estimate the effective thermal conductivity, we use the Hamilton-Crosser model \cite{34,39,40}, which is
\begin{eqnarray}
K_d = K_{air}\frac{K_{Fe}+2K_{air}+2\Phi\left(K_{Fe}-K_{air}\right)}{K_{Fe}+2K_{air}-\Phi\left(K_{Fe}-K_{air}\right)},
\label{eq:3}
\end{eqnarray}
where $K_{air} = 0.03$ W/m/K \cite{1} and $K_{Fe} = 80$ W/m/K \cite{20}. The Hamilton-Crosser model has been found to provide accurate quantitative descriptions of effective thermal conductivity \cite{39}. Meanwhile, the effective specific heat capacity and mass density are given by \cite{27}
\begin{eqnarray}
C_p &=& \frac{ C_{p,air}\rho_{air}(1-\Phi)+C_{p,Fe}\rho_{Fe}\Phi}{\rho_{air}(1-\Phi)+\rho_{Fe}\Phi} ,\nonumber\\
\rho &=&  \rho_{air}(1-\Phi) + \rho_{Fe}\Phi,
\label{eq:4}
\end{eqnarray}
where $\rho_{air}=1.293$ kg/\ce{m^3} and $\rho_{Fe}=7874$ kg/\ce{m^3} are the mass density of air and iron, respectively, and $C_{p,air}=1000$ \ce{J/kg/K} and $C_{p,Fe}=480$ \ce{J/kg/K} are the specific heat capacity of air and iron, respectively, \cite{20}.

\begin{figure}[htp]
\includegraphics[width=9cm]{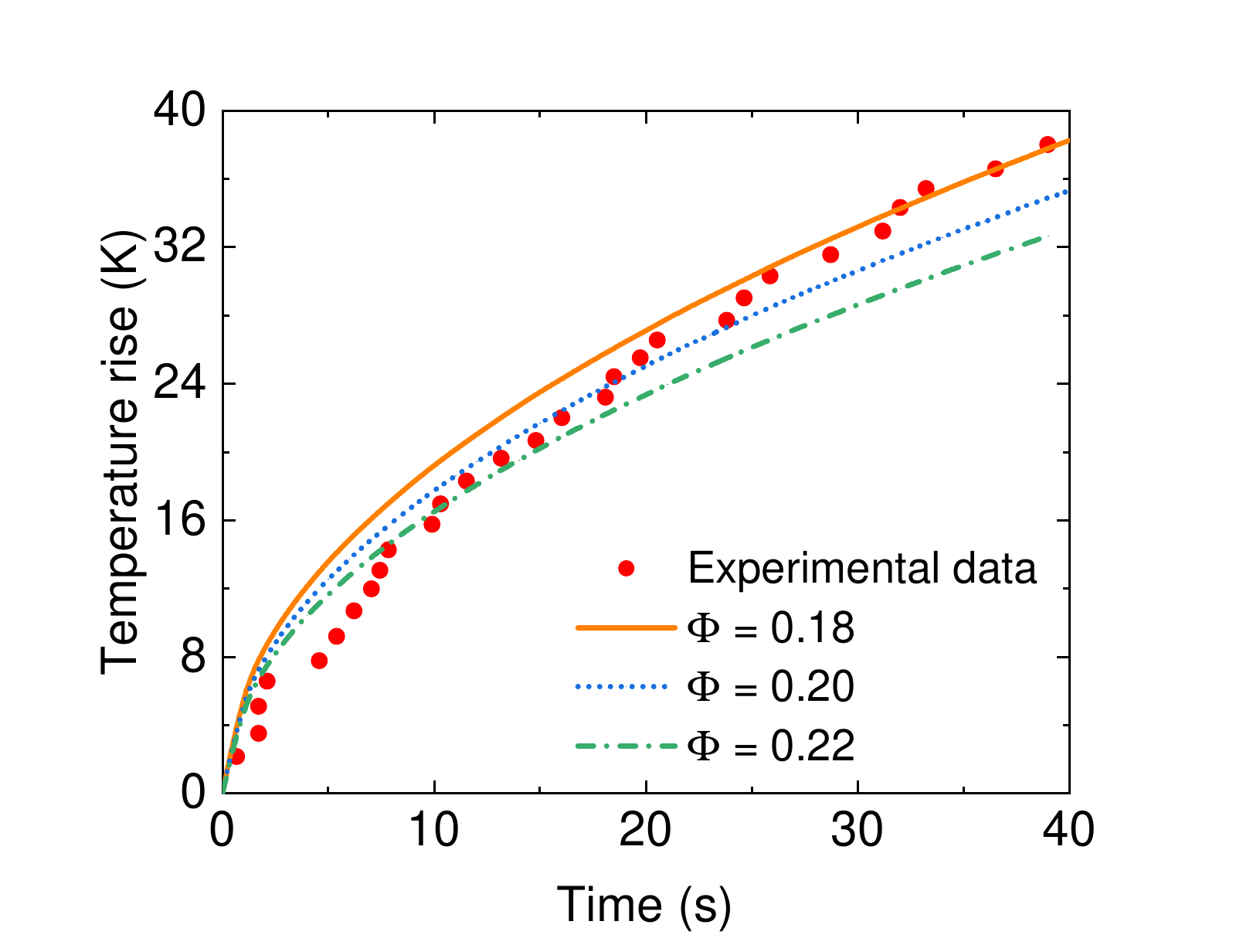}
\caption{\label{fig:8}(Color online) Experimental data \cite{1} and theoretical calculations for the temperature rise of FeCoNi and Fe nanoparticles with an average radius $R=30$ nm at $r=z=0$ as a function of time under solar light illumination, respectively. Numerical results are calculated using Eqs. (\ref{eq:2}), (\ref{eq:3}), and (\ref{eq:4}). The values of $\Phi$ are adjusted to obtain good fits between our model and the experimental data.}
\end{figure}

By using the absorption cross section of the Fe nanoparticle ($Q_{abs}$ in Fig. \ref{fig:5}) calculated by pyGDM simulations associated with the AM1.5 solar spectrum \cite{59} and the effective thermal conductivity, specific heat capacity, and mass density derived from Eqs. (\ref{eq:3}) and (\ref{eq:4}), respectively, we calculate the time-dependent temperature increase at the Fe nanoparticle surface hotspot (Fig. \ref{fig:8}) under solar irradiation for various nanoparticle volume fractions. Our approach, consistent with previous simulations \cite{22, 23, 24}, assumes that the thermal conductivity, specific heat capacity, and mass density of the metal remain nearly constant over the considered temperature range. This assumption is relatively reasonable as these properties exhibit minimal thermal sensitivity near room temperature. Figure \ref{fig:8} compares our theoretical predictions for the iron nanoparticle substrate with the experimental photothermal data for the FeCoNi nanoparticle system reported in Ref. \cite{1}. By adjusting the volume fraction of iron nanoparticles, we achieve good quantitative agreement with the experimental data. This agreement validates the ability of our model to accurately capture the essential physics of the photothermal heating process. Furthermore, it suggests that the optical and photothermal properties of FeCoNi-based alloys can be effectively approximated by those of pure iron nanoparticles. Our finding potentially simplifies future investigations of these promising materials. Although increasing the volume fraction of Fe nanoparticles leads to greater optical energy absorption, the simultaneous increase in effective thermal conductivity facilitates heat dissipation and counteracts the expected temperature rise.

\section{Conclusions}
In this study, we have introduced a comprehensive theoretical approach to investigate optical and photothermal properties of FeCoNi-based HEA nanoparticles. We determined the effective dielectric function of alloys by two different models: (1) the random mixing model of alloy constituents and (2) the Maxwell-Garnett model for effective medium approximation. These dielectric functions are then incorporated into pyGDM simulations to calculate absorption spectra for both pure metal and alloy nanoparticles. By comparing the simulated spectra with experimental data, we selected the most suitable dielectric datasets for each constituent element in the high-entropy alloys and found that using the random mixing model in pyGDM simulation provides more accurate and reasonable results than that based on the Maxwell-Garnett model. Thus, only pyGDM simulation associated with the random mixing model is used to further compute the absorption spectra of FeCoNi-based alloy nanoparticles. Our numerical results revealed that optical properties of FeCoNi-based nanoparticles are similar to those of pure iron nanoparticles. This suggests that the thermal responses of iron nanoparticles can be exploited to describe those of FeCoNi-based alloy nanoparticles. To validate this assumption, we used another theoretical model based on solving the heat energy balance equation and using the absorption prediction to calculate the time-dependent temperature rise of iron-nanoparticles-based substrate under solar irradiation. The quantitative agreement between our theoretical temperature rise of Fe nanoparticles and the experimental photothermal data for FeCoNi nanoparticles validates our assumptions. This correspondence allows for a simplified modeling approach, wherein iron nanoparticles can effectively represent the complex optical and photothermal behavior of FeCoNi-based HEAs. Overall, our work not only simplifies the theoretical modeling of these complex alloys but also provides valuable insights into the optical and photothermal behaviors of high-entropy alloy nanoparticles.

\begin{acknowledgments}
Do T. Nga acknowledges support from International Centre of Physics at the Institute of Physics, VAST, under Grant No. ICP.2024.05.
\end{acknowledgments}
\section*{Conflicts of interest}
There are no conflicts to declare.

\section*{Data Availability}
The data that supports the findings of this study are available within the article.

\end{document}